\documentclass[preprint,prb,showkeys,floatfix]{revtex4}
\usepackage{graphics}
\usepackage{setspace}
\usepackage{graphicx}
\usepackage{amssymb}

\begin{document}

\title{Quantum control and manipulation of donor electrons in Si-based quantum computing}
\author{Mar\'{\i}a Jos\'e Calder\'on}
\affiliation{Instituto de Ciencia de Materiales de Madrid (CSIC), Cantoblanco, 28049 Madrid (Spain)}
\author{Andr\'e Saraiva}
\affiliation{Instituto de F\'{\i}sica, Universidade Federal do Rio de
Janeiro, Caixa Postal 68528, 21941-972 Rio de Janeiro, Brazil}
\author{Belita Koiller}
\affiliation{Instituto de F\'{\i}sica, Universidade Federal do Rio de
Janeiro, Caixa Postal 68528, 21941-972 Rio de Janeiro, Brazil}
\author{Sankar {Das Sarma}}
\affiliation{Condensed Matter Theory Center, Department of Physics,
University of Maryland, College Park, MD 20742-4111}

\date{\today}

\begin{abstract}
Doped Si is a promising candidate for quantum computing due to its scalability properties, long spin coherence times, and the astonishing progress on Si technology and miniaturization in the last few decades. This proposal for a quantum computer ultimately relies on the quantum control of electrons bound to donors near a Si/barrier (e.g. SiO$_2$) interface. We address here several important issues and define critical parameters that establish the conditions that allow the manipulation of donor electrons in Si by means of external electric and magnetic fields.  
\end{abstract}

\keywords{silicon devices,quantum computing}

\maketitle
%\doublespacing
\section{Introduction}
In the last few decades, the density of transistors in a chip has been consistently duplicated every two years, as given by Moore's law. This demand is pushing the semiconductor devices fabrication (in particular, for Si) towards the atomic limit in which the number of dopants is so small that their exact position and distribution affects the performance of a transistor.~\cite{shanida05} There is therefore a great interest in achieving an atomic control over the position of impurities in Si.~\cite{ruessNANO04,ruessSMALL07} This interest has been reinforced by the proposal of using shallow donors (typically P or As) as spin qubits for a Si-based quantum computer,~\cite{kane98,kane00} and related ones that followed,~\cite{vrijen00} where qubit operations are performed by manipulating the donor electron with an external electric field provided by local gates. Spin is a natural candidate for a qubit, particularly in Si where very long spin coherence times have been measured (at least $\sim 1$ ms in natural Si).~\cite{tyryshkin03,sousa03,witzel05,tyryshkin06} Spin coherence time has to be orders of magnitude longer than the time required for operations, and enough to allow for quantum error correction.\cite{shorPRA95}

\begin{figure}
\begin{center}
\resizebox{60mm}{!}{\includegraphics{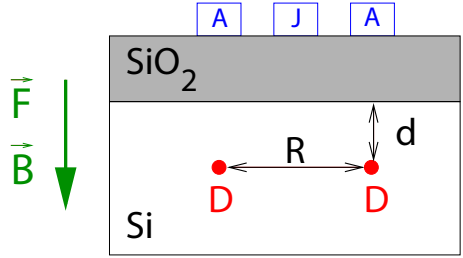}}
\caption{\label{fig:QCproposal} Basic scheme of the doped Si-based proposal for quantum computing.~\cite{kane98} It consists of an array of donors in Si separated by a distance $R$ and located at a distance $d$ from a SiO$_2$ barrier. The donor electrons are manipulated by applying electric fields by local gates. The A gates control the relative position of the electron with respect to the donor affecting the hyperfine interaction between their spins (allowing 1-qubit operations). The J gates control the overlap between neighboring electrons, allowing to  perform the  exchange pulse involved in 2-qubit operations.
}
\end{center}
\end{figure}

In the doped Si-based quantum computer,~\cite{kane98,kane00} schematically shown in Fig.~\ref{fig:QCproposal}, single-qubit operations (rotations of the spin state on the Bloch sphere) are accomplished with an AC magnetic field in resonance with the level splitting of the hyperfine coupled nuclear spin-electron spin system. The hyperfine coupling is proportional to the probability density of the electron wave-function on the donor. Therefore, the on and off resonance condition is controlled by applying electric fields to manipulate the electron wave-function. Two-qubit operations could in principle be performed by letting electrons in neighboring donors interact via a transient exchange coupling during a specified period of time.~\cite{loss98} Combinations of these exchange mediated operations and single spin rotations configures a controlled-NOT (CNOT) gate.~\cite{nielsenchuang}  With the one qubit rotations and the CNOT gate any operation on $n$ qubits may be performed.~\cite{barenco95,nielsenchuang}

\begin{figure}
\begin{center}
\resizebox{70mm}{!}{\includegraphics{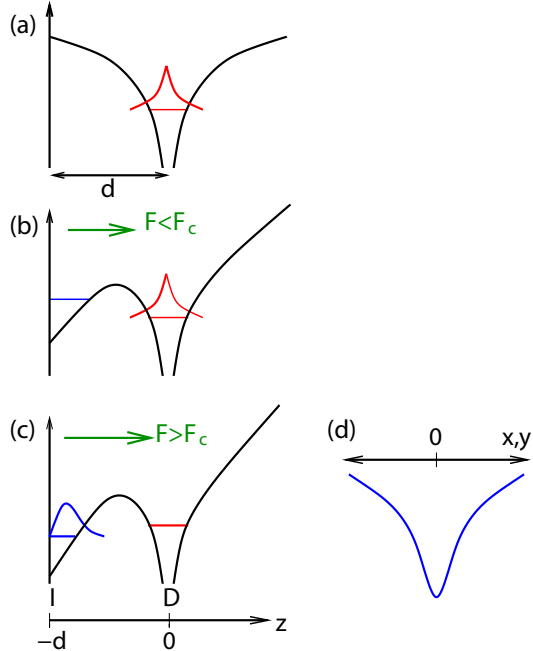}}
\caption{\label{fig:scheme-D-I} (a) If no electric field is applied the electron only sees the donor potential and is bound to it. (b) When an electric field $F$, perpendicular to the interface, is applied, a triangular shaped potential forms at the interface. For a certain value of $F$ a bound state exists at the interface. (c) When $F  \sim F_c$ the states at the interface and at the donor are degenerate and tunneling between donor and interface may occur. (d) Donor potential at the interface along the xy-plane $V(x)={{2(Q-1)}\over{\sqrt{x^2+d^2}}}$. This attractive potential causes the electron not to spread along the interface when it is ionized from the donor. 
}
\end{center}
\end{figure}

We have considered the problem of a donor a distance $d$ from a Si/SiO$_2$ (001) interface and studied the manipulation of the donor electron by electric and magnetic fields. Our study applies for cases when $d$ is large enough so that, under an applied field, abrupt ionization (via tunneling) takes place.~\cite{martins04} When no external fields are applied, the only attractive potential felt by the electron is the donor Coulomb potential and, at the low temperatures relevant here, the electron remains bound to the donor [see Fig.~\ref{fig:scheme-D-I}(a)]. Under an electric field ($F$) applied perpendicular to the interface, a triangular potential well is formed at the interface, as shown in Fig.~\ref{fig:scheme-D-I}(b). For fields larger than a certain minimum value, bound states exist at the triangular well. When $F$ reaches a characteristic strength $F_c$ such that the bound state at the interface is degenerate in energy with the bound state at the donor, tunelling between the two wells (the donor Coulomb potential, and the triangular well at the interface) is possible. In this way, we can reversibly manipulate the position of the electron between the donor and the interface by oscillating the electric field around $F_c$ [namely, going back and forth from situation in Fig.~\ref{fig:scheme-D-I}(b) to situation in  Fig.~\ref{fig:scheme-D-I}(c)]. Note that the electron, when at the interface, may still remain bound to the donor along the xy-plane, and hence, the electron may be confined in all three space directions upon certain conditions that will be described below. We simplify the problem by assuming that the SiO$_2$ barrier is infinite.

The outline of this paper is the following. In Sec.~\ref{sec:model} we describe our model and discuss the peculiarities of the conduction band of Si, that has a multivalley structure. We also describe our basis set consisting on the lowest uncoupled donor and interface states. In particular, we point out the conditions that guarantee the confinement in all 3 directions of space of the electron at the interface. In Sec.~\ref{sec:manipulation} we discuss the conditions that allow the manipulation of the donor electron to be possible, in particular:~\cite{calderonPRL06,calderon07} (i) we calculate the value of the characteristic electric field at which shuttling of the electron between donor and interface may occur;  and(ii) we estimate the tunneling time of this process. We also briefly discuss the effect of a magnetic field applied parallel to the electric field.~\cite{calderon062} In Sec.~\ref{sec:exchange} we discuss how the 2-qubit operations could be performed with the electrons at the interface state~\cite{calderon07} and show new results for the calculation of the exchange calculated with an improved Heitler-London method.~\cite{saraiva07}
We finish with a discussion and conclusions in Sec.~\ref{sec:discuss}.

\section{Model}
\label{sec:model}
The wavefunction of a donor electron in a semiconductor can be written as an expansion in terms of the Bloch waves close to the bottom of the conduction band.~\cite{kohn} The conduction band of Si has six minima (valleys) located in the $\langle 100 \rangle$ directions at a distance $k_0=2\pi \,0.85/a_{Si}$ ($a_{Si}=5.4$ \AA $\,\,$ the lattice parameter of Si) from the $\Gamma$ point. Therefore, combinations of Bloch waves from the 6 valleys have to be considered, and the donor ground state wave-function would be 6-fold degenerate. This degeneracy is lifted  when the coupling between Bloch waves from different valleys due to the singular donor coulomb potential is taken into account (the so-called valley-orbit coupling). The resulting ground state is non-degenerate and it is a symmetric combination of the six valleys (symmetry A$_1$).~\cite{kohn} Within the effective mass approximation, only the Bloch functions at the positions of the conduction band minima are involved, and the ground state of the electron at the donor is written~\cite{kohn55}
\begin{equation}
\Psi_D^{\rm gs}= {{1}\over{\sqrt{6}}} \sum_{\mu} F_D^{\mu} ({\bf r}) \phi_{\mu} ({\bf r},{\bf r}_D) \,,
\label{eq:psid}
\end{equation}
where $F_D^\mu ({\bf r})$ are envelope functions ($\mu$ = +x,-x,+y,{\mbox -y,+z,-z}), and
$\phi_{\mu} ({\bf r},{\bf r}_D)= \Psi_{\rm Bloch}^{{\bf k}_{\mu}} e^{-i {\bf k}_{\mu}\cdot {\bf r}_D} =u_{\mu} ({\bf r}) e^{i {\bf k}_{\mu}\cdot ({\bf r}-{\bf r}_D)}$. Here, ${\bf r}_D$ is a reference site, which represents a pinning site for the Bloch waves in the superposition state. In the case of a single Bloch state, ${\bf r}_D$ contributes with an irrelevant phase, but for superposition states it leads to interference effects. For an isolated impurity, ${\bf r}_D$ is naturally chosen to be the position of the donor ($r_D=0$).

For the description of the interface state it has to be taken into account that the (001) interface breaks the degeneracy of the 6 valleys, making the x, -x, y, and -y valleys much higher in energy than the z and -z valleys.~\cite{kane00PRB} (A tensile strain has the same effect on the valley degeneracy).~\cite{koiller02PRB} Therefore, the ground state at the interface only includes the Bloch states from the z and -z valleys. We write it
\begin{equation}
\Psi_I^{\rm gs}={{1}\over{\sqrt{2}}} F_I  ({\bf r}) \left[ \phi_{z} ({\bf r},z_I+d) \pm \phi_{-z}({\bf r},z_I+d)\right]
\end{equation}
where the envelope function $F_I$ is obtained variationally assuming the form:
\begin{equation}
F_I=  {{\alpha^{\frac{5}{2}}}\over{2\sqrt{3}}} (z+d)^{2}\, e^{-{\alpha (z+d)}/{2}}\times{{\beta}\over{\sqrt{\pi}}}\, e^{{-\beta^2 \rho^2}/{2}}
\label{eq:FI}
\end{equation}
with $\alpha$ and $\beta$ taken as variational parameters, which are the same as for the single valley approximation.~\cite{calderonPRL06} The parameter $1/\alpha$ is related to the width of the wave-function at the interface in the z-direction while $1/\beta$ gives the width of the wavefunction along the xy plane.~\cite{calderonPRL06,calderon07} For the isolated interface, we assume the pinning site $z_I$ is exactly at the interface.

The value of $1/\beta$ depends on how far the donor is from the interface. When the electron is at the interface it still feels the attractive potential of both the donor and its image [see Fig.~\ref{fig:scheme-D-I} (d)]. Typical values of $1/\beta$ range from $6$ nm for $d=6$ nm to $22$ nm for $d=40$ nm. If we want to be able to shuttle the electron between the interface and the donor reversibly, we need the electron to be confined at the interface, and not to spread forming a two-dimensional gas. This gives us a limit for the maximum planar density of around $10^{10}$ cm$^{-2}$.~\cite{calderonPRL06,calderon07} 

The fact that the 6-valley degeneracy is broken at the interface  plus the fact that the smooth potential between the donor and the interface does not couple valleys in different directions, allows us, in a first approach to the problem, to consider only the z and -z valleys for the state at the donor. This simplification would be exact for  strained Si. Therefore, it makes sense to use
\begin{equation}
\Psi_D^{\rm gs}={{1}\over{\sqrt{2}}} F_D  ({\bf r}) \left[ \phi_{z} ({\bf r},z_D) \pm \phi_{-z}  ({\bf r},z_D)\right]
\end{equation}
with
\begin{equation}
F_D \propto (z+d)e^{-\sqrt{\rho^2 / a^2+z^2/b^2}}~,
\label{eq:FD}
\end{equation}
the donor envelope function centered at ${\bf r}=0$. This is a hydrogen-like envelope with an anisotropic shape that arises due to the conduction band effective mass anisotropy in Si, multiplied by the factor $(z+d)$ that guarantees that the wavefunction fulfills the boundary condition for an infinite barrier at $z=-d$. We are interested here in the situation where the donor is relatively far from the interface, i.e, the ground state wavefunctions in each well do not overlap significantly, and in this range the variationally determined Bohr radii $a$ and $b$ are the same as for a donor in bulk ($a=2.365$ nm and $b=1.36$ nm).~\cite{calderon07}

Note that, for each isolated well, the value of the splitting between the symmetric and antisymmetric combinations of the Bloch waves and which one is the ground state depend on the value of the pinning positions $z_D$ or $z_I$, as illustrated in Fig.~\ref{fig:vo}.

\begin{figure}
\begin{center}
\resizebox{70mm}{!}{\includegraphics{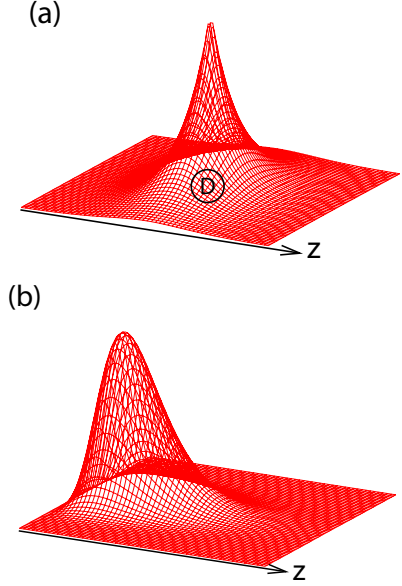}}
\caption{\label{fig:envelopes} Typical plots of the shapes of the envelope functions at the donor (a) given by $F_D$ Eq.~(\ref{eq:FD}), and at the interface (b) given by $F_I$ Eq.~(\ref{eq:FI}). $F_I$ is much more extended that $F_D$ in the xy-plane ($1/\beta \sim 10-20$ nm, depending on $d$, ~\cite{calderonPRL06,calderon07} while the Bohr radii $a=2.365$ nm).
}
\end{center}
\end{figure}

\begin{figure}
\begin{center}
\resizebox{70mm}{!}{\includegraphics{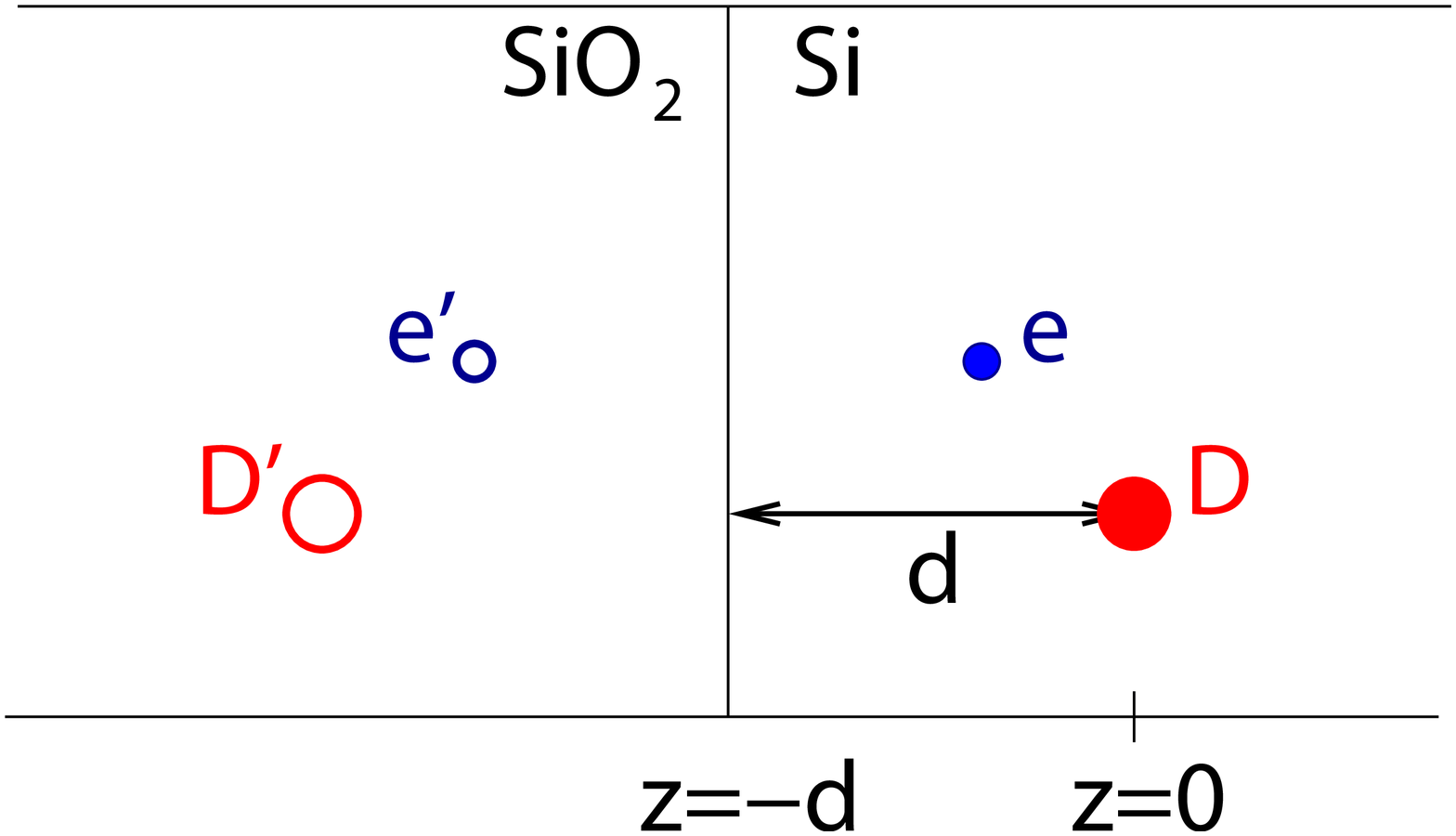}}
\caption{\label{fig:images} We consider Si and SiO$_2$ to be two semi-infinite layers. We include the image charges D' and e'. For our particular value of $Q=-0.5$  we see that the sign of the images is the same as that of the original charges. Therefore, the donor image potential is attractive and the electron image potential is repulsive [see the expressions for $V_D^{\rm image}$ Eq.~(\ref{eq:vdimage}) and $V_e^{\rm image}$ Eq.~(\ref{eq:veimage})].
}
\end{center}
\end{figure}

\begin{figure}
\begin{center}
\resizebox{80mm}{!}{\includegraphics{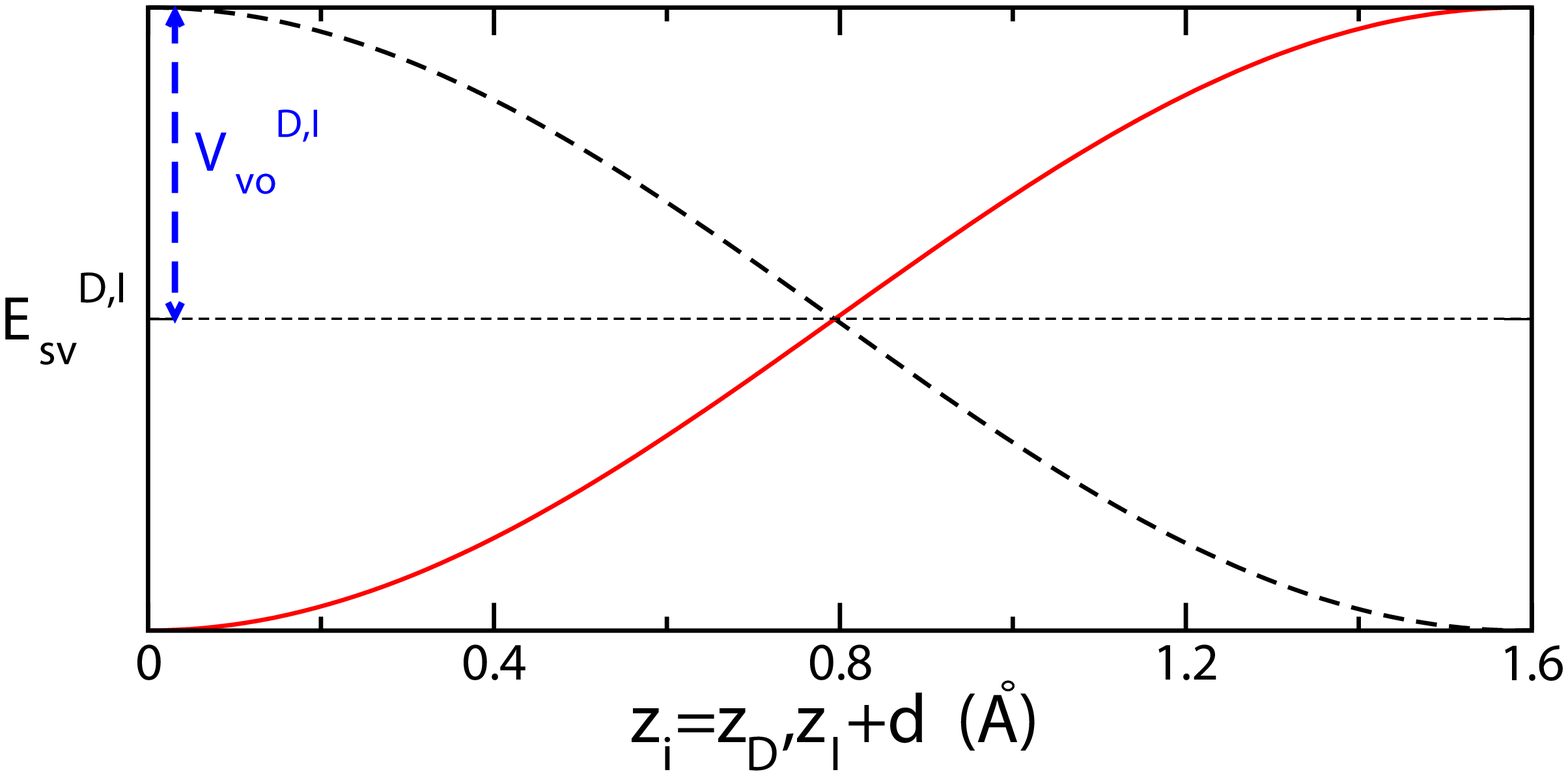}}
\caption{\label{fig:vo} The expectation values for the energies corresponding to the symmetric (solid line) and antisymmetric (dashed line) combinations of the z and -z Bloch waves depend on the value of the pinning site $z_i$, varying as  $\cos(2 k_0 z_i)$. For $z_i=0$ the ground state is the symmetric combination and the splitting between the two combinations is maximum and equal to $2 V_{vo}^{i}$. But for very small displacements of the pinning site (of less that $1$  \AA) the splitting changes enormously. The dotted horizontal line corresponds to the ground state energy within the single valley approximation.~\cite{calderonPRL06,calderon07}
}
\end{center}
\end{figure}

Let us write now the full Hamiltonian for a donor (at $z=0$) in Si a distance $d$ from a (001) interface with SiO$_2$. The boundary problem between the two semi-infinite dielectrics is considered by including the image charges for the donor and for the electron (see Fig.~\ref{fig:images}). A uniform electric field, perpendicular to the interface, is also included. The resulting Hamiltonian is,~\cite{calderonPRL06,calderon07}
in rescaled atomic units  $a^*={{\hbar^2\epsilon_{Si}}/{m_\perp e^2}} =
 3.157$~nm and $Ry^*={{m_\perp e^4}/{2\hbar^2\epsilon_{Si}^2}}=
 19.98$~meV,
\begin{equation}
H = T+\kappa e F z -{{2}\over{r}}+V_D^{\rm image}+V_e^{\rm image}
+H_{vo}~,
\label{eq:h-effunits}
\end{equation}
where
$T=- {{\partial^2}\over{\partial x^2}} - {{\partial^2}\over{\partial
 y^2}}- \gamma {{\partial^2}\over{\partial
 z^2}}$, $\gamma=m_\perp/ m_\|$, $m_\perp=0.191m$, $m_\|=0.916$, $\kappa=3.89 \times 10^{-7} \epsilon_{Si}^3\left({{m}/{m_\perp}}\right)^2$ cm/kV, and the electric field $F$ is
given in kV/cm. The next terms are the donor image
\begin{equation}
V_D^{\rm image}={{2Q}\over{\sqrt{\rho^2+(z+2d)^2}}} \, ,
\label{eq:vdimage}
\end{equation}
and the electron image
\begin{equation}
V_e^{\rm image}= -{{Q}\over{2(z+d)}} \, .
\label{eq:veimage}
\end{equation}
$Q$ is a function of the dielectric constants $Q={{(\epsilon_{\rm SiO_{2}}-\epsilon_{\rm Si})}/{(\epsilon_{\rm
 SiO_{2}}+\epsilon_{\rm Si})}}$, with
$\epsilon_{\rm Si}=11.4$ and  $\epsilon_{\rm SiO_{2}}=3.8$. With this $Q$ we get that the donor image potential is attractive while the electron image potential is repulsive.

The last term in the Hamiltonian Eq.~(\ref{eq:h-effunits}) describes valley-orbit effects, namely the coupling
between different valleys due to the singular nature of both the donor
$(D)$ and the interface $(I)$ potentials. These couplings are
quantified by the parameters $V_{vo}^D$ and $V_{vo}^I$. $V_{vo}^D$ is known from the splitting of the 1S manyfold of the isolated donor spectrum $V_{vo}^D=-1.5$ meV. We use this value all throughout.  $V_{vo}^I$ is not precisely known, although estimated to be in the order of $1$ meV,~\cite{ando82} and probably dependent on the interface quality.~\cite{takashina06,saraiva09} We consider it as a parameter ranging from $0$ to $-10$ meV. The results described below for a finite value of the interface valley orbit coupling do not depend qualitatively on the particular value or sign of $V_{vo}^I$ for  $|V_{vo}^I| \gtrsim 0.02$ meV.

In our two-valley formalism, assuming that $d$ is large enough so that no strong hybridization occurs between the donor and interface states, the problem may be restricted to the basis set of the lowest uncoupled donor and interface states, namely $\{F_D \Psi_{\rm Bloch}^{k_z} e^{-i k_0 z_D}$, $F_D \Psi_{\rm Bloch}^{-k_z} e^{i k_0 z_D}$, $F_I \Psi_{\rm Bloch}^{k_z} e^{-i k_0 z_I}$, $F_I \Psi_{\rm Bloch}^{-k_z} e^{i k_0 z_I}\}$, leading to the Hamiltonian matrix

\begin{widetext}
\begin{equation}
{\mathbf H} =\left( \begin{array}{cccc}
 E_{D} & V_{vo}^D \cos(2 k_0 z_D) & H_{ID} e^{ i k_0 (z_D-z_I)} & 0 \\
 V_{vo}^D \cos(2 k_0 z_D) &E_{D} &  0 & H_{ID} e^{ -i k_0 (z_D-z_I)}\\
 H_{ID} e^{- i k_0 (z_D-z_I)} & 0 &E_I & V_{vo}^I \cos[2 k_0 (z_I+d)]\\
0 & H_{ID} e^{ i k_0 (z_D-z_I)}  & V_{vo}^I \cos(2 k_0 (z_I+d)) &E_I
 \end{array} \right)~,
\label{eq:2by2matrix}
\end{equation}
\end{widetext}
where $E_D=\langle F_D|H|F_D \rangle$, $E_I=\langle F_I|H|F_I \rangle$, and $H_{ID}=\langle F_I|H|F_D \rangle$ are the same as the single-valley matrix elements.~\cite{calderonPRL06}

\section{Manipulation of a single donor electron close to a Si/SiO$_2$ interface}
\label{sec:manipulation}

We solve the Hamiltonian Eq.~(\ref{eq:2by2matrix}) for different $d$'s and a range of electric fields $F$. At each point, the pinning sites $z_I$ and $z_D$ are determined variationally such that the total ground state energy is minimized. For $F<<F_c$, the electron is at the donor and the pinning is on the donor, to be precise, $abs[\cos(2 k_0 z_D)]=1$ while $abs[\cos(2 k_0 (z_I+d))] < 1$. In the opposite limit, $F>>F_c$, the electron is confined at the interface and so is the pinning, namely, $abs[\cos(2 k_0 (z_I+d))] = 1$ while $abs[\cos(2 k_0 z_D)]<1$.~\cite{calderon07}

\begin{figure}
\begin{center}
\resizebox{100mm}{!}{\includegraphics{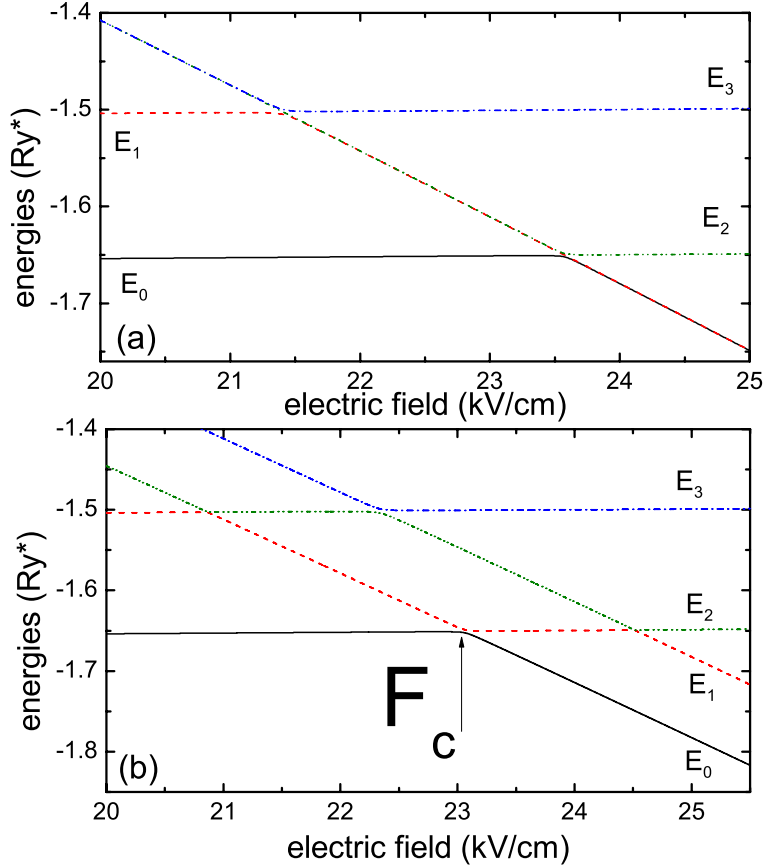}}
\caption{\label{fig:antic} (a) Eigenvalues of the Hamiltonian in Eq. (\ref{eq:2by2matrix}) for $d=5.99 a^*$ and $V_{vo}^I=0$. There are two anticrossings involving three levels each (the intermediate level remains uncoupled to the other two, which display the anticrossing).  (b) Eigenvalues of the problem for $d=5.94 a^*$ and $V_{vo}^I=-1$ meV. The four levels show anticrossings in pairs. The relevant anticrossing for the determination of $F_c$ and the tunneling times (or, equivalently, gaps) is the one involving the ground state. The gaps at anticrossing are small at the energy scale of the figure.
}
\end{center}
\end{figure}

The variational solution of the Hamiltonian Eq.~(\ref{eq:2by2matrix}) leads to four eigenvalues that, as a function of the electric field, show certain level anticrossings.  As an illustration, two typical cases  are shown in Fig.~\ref{fig:antic}.~\cite{multivalley} When $V_{vo}^{I}=0$, the two states at the interface are degenerate and only two anticrossings, involving three levels each, occur. For the more general case of 
$V_{vo}^{I} \ne 0$, there are typically four anticrossings, involving two levels each. 
For particular values of $d$, corresponding to $\cos(k_0 d)=0$, the coupling between levels is so strong that one anticrossing can involve all four levels.  

\begin{figure}
\begin{center}
\resizebox{80mm}{!}{\includegraphics{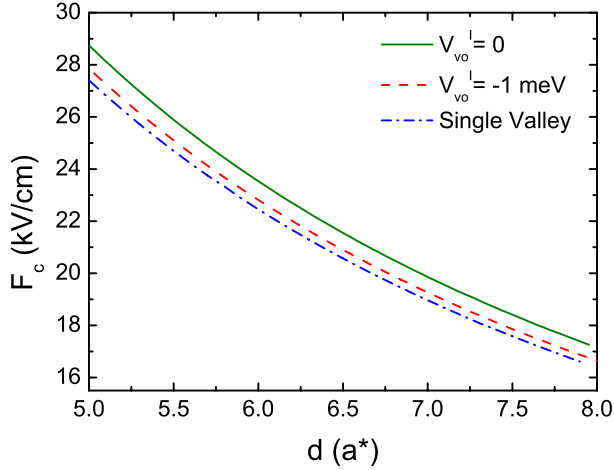}}
\caption{\label{fig:Fc} Characteristic field at which tunneling between the donor and the interface state may occur. The three curves (corresponding to single valley approximation, 2-valley with $V_{vo}^I=0$, and 2-valley with $V_{vo}^I=-1$ meV) have the same qualitative behavior.
}
\end{center}
\end{figure}

To determine the characteristic field $F_c$ at which the 'shuttling' of the electron between donor and interface may occur and the times involved in the process, we need to look at the anticrossing between the lowest eigenvalues. The value of the electric field at which the anticrossing happens is $F_c$ and the gap between levels at anticrossing gives an estimate for the tunneling time $\tau\sim\hbar/E_{gap}$. In Fig.~\ref{fig:Fc} we show the value of $F_c$ versus $d$ for the single valley approximation, the 2-valley model with $V_{vo}^I=0$, and $V_{vo}^I=-1$ meV. The three curves show the same qualitative shape. The small shift between them arises due to the changes produced in the ground state energies when the valley orbit coupling is included (this change is illustrated, with respect to the single valley energy, in Fig.~\ref{fig:vo}). Therefore, the multivalley structure of the conduction band of Si does not significantly affect the value of $F_c$. This was, in fact, expected from previous tight-binding results that took into account the six valleys.~\cite{martins04} Very recent experiments on a Si FinFET with a single donor~\cite{lansbergen08} show the same kind of behavior for $F_c$ versus $d$ that we find here.  

\begin{figure}
\begin{center}
\resizebox{80mm}{!}{\includegraphics{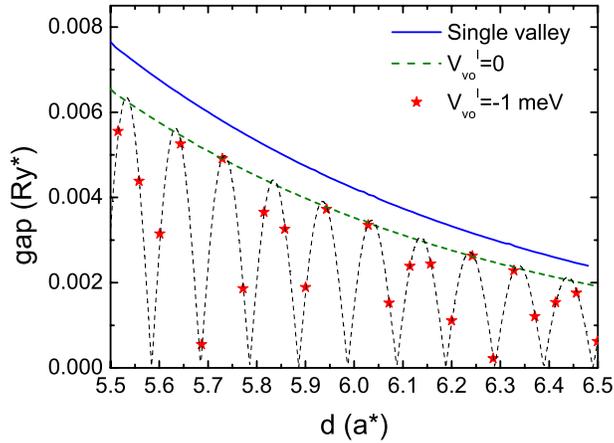}}
\caption{\label{fig:gap} Gap (related to tunneling times as  $\tau\sim\hbar/E_{gap}$) versus $d$ for single valley approximation, $V_{vo}^I=0$, and $V_{vo}^I=-1$ meV. The first two curves smoothly decay with $d$. However, the inclusion of a finite valley-orbit coupling at the interface leads to oscillations on the gap that go as $abs[\cos(k_0 d)]$. The stars correspond to values of $d$ given by monolayer steps of $1.35$ \AA.   
}
\end{center}
\end{figure}

Now we look at the value of the gap between levels calculated at anticrossing: The results are shown in Fig.~\ref{fig:gap}. For the 2-valley model with degenerate states at the interface ($V_{vo}^I=0$) the result is qualitatively similar to the smooth decrease already observed in the single valley approximation.~\cite{calderonPRL06} This is due to the fact that only one of the degenerate levels at the interface is coupled to the lowest level of the donor, leading to an effectively single valley result. However, a qualitative striking difference is observed when a finite valley orbit coupling is present at the interface: The gap versus $d$ shows oscillations that go as $abs[\cos(k_0 d)]$, namely, the oscillations are not commensurate with the lattice. The tunneling times we get within the single valley approximation range from subpicosecond to nanoseconds depending on $d$ (larger $d$ correspond to longer vales of $\tau$). The times corresponding to the multivalley result can be much larger for $d$'s such that $\cos(k_0 d)\sim 0$, but the statistical weight of such points (see the stars in Fig.~\ref{fig:gap}) is relatively small. Therefore, in an ion implanted sample, where the control on the donor  depth positioning is limited, most donors would have tunneling times comparable to the single valley results, while others would take much longer.

We have also analyzed the effect of a magnetic field applied parallel to the electric field.~\cite{CKDmag,calderon07} The magnetic field shrinks the electron wavefunction in the direction parallel to the interface. The effect is much stronger for less confined wavefunctions. Consequently, the wavefunction at the interface is much more affected than at the donor.  The increase in confinement is concomitant with an increase in energy, which is hence significant at the interface while very small at the donor for magnetic fields on the order of a few Tesla. This magnetic field induced shift in energy 
has an interesting consequence: Starting from an electric field just above $F_c$ (so the electron is at the interface), the application of a magnetic field can push the electron back to the donor.~\cite{calderon07} The manipulation of electrons using simultaneously electric and magnetic fields can help us distinguish donor electron from other moving charges in real systems.~\cite{kenton06}

\section{Exchange between electrons at the interface}
\label{sec:exchange}

We have discussed above how the multivalley nature of the conduction band of Si produces oscillations in the tunneling time involved in the shuttling of an electron between donor and interface. Similar oscillations have been found in the exchange interaction between electrons in neighboring donors as a function of the distance $R$ between them.~\cite{koiller02PRL}   

\begin{figure}
\begin{center}
\resizebox{65mm}{!}{\includegraphics{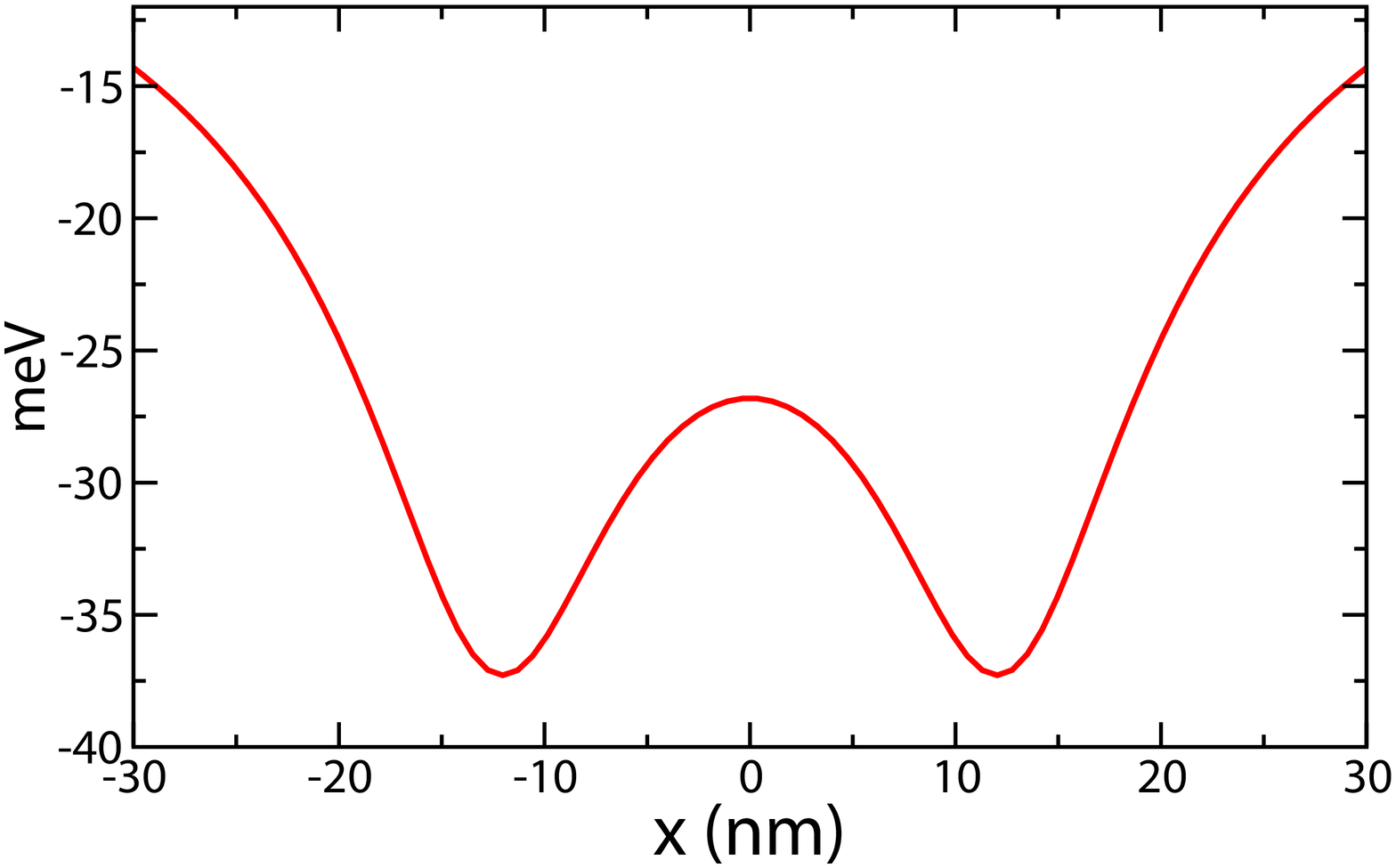}}
\caption{\label{fig:int-double-well} Donor pair potential at the interface, along the axis connecting the wells minima as given by Eq.~(\ref{eq:int-double-well}) for $R=28$ nm and $d=6.3$ nm.
}
\end{center}
\end{figure}

Since the donor electron is much less confined at the interface than at the donor well, the transient exchange interaction between donors very distant from each other in the bulk, where they do not interact,   could be performed by pulling (with an electric field) the two corresponding electrons towards the interface, where they would interact. Besides, if the two neighboring donors are at the same distance $d$ from the interface, we do not  expect to get oscillations on the exchange versus interdonor distance $R$. The 2D interface potential formed by two donors separated a distance $R$ from one another and at the same depth $d$ from the interface has the double-well shape, depicted in Fig.~\ref{fig:int-double-well} along the  axis ($x$) connecting the well minima, as given by the expression
\begin{equation}
V(x)={{2(Q-1)}\over{\sqrt{(x-R/2)^2+d^2}}} +{{2(Q-1)}\over{\sqrt{(x+R/2)^2+d^2}}}.
\label{eq:int-double-well}
\end{equation}

\begin{figure}
\begin{center}
\resizebox{80mm}{!}{\includegraphics{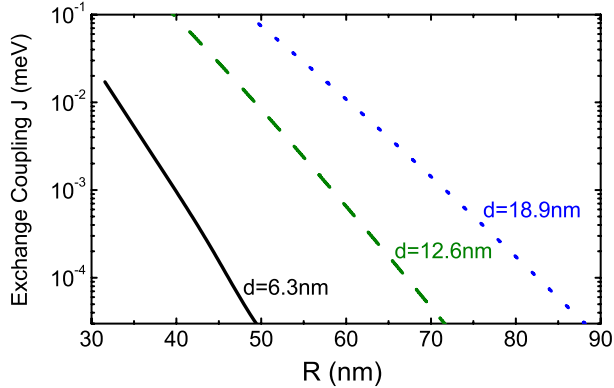}}
\caption{\label{fig:exchange} Exchange calculated within the Heitler-London approximation using a matched variational wavefunction that correctly includes exponential decaying tails. 
}
\end{center}
\end{figure}

The distance $R$ between donors has to be large enough so that the single electron wavefunction is below the barrier between the wells, to guarantee that the electron shuttling between donor and interface is reversible. In Fig.~\ref{fig:exchange} we show the exchange calculated for donor pairs at three different distances from the interface as a function of the interdonor distance. The calculations were performed in 2D, using the Heitler-London approximation with an improved hybrid variational wave-function which is gaussian in the center, where the parabolic approximation for the potential is valid and has an exponential decay at long distances of the well minimum, where the potential saturates and deviates significantly from the parabolic behavior,~\cite{saraiva07}

\begin{equation}
\phi_{MV}(\rho) = \left\{ \begin{array}{ll}
 A_1 \exp{\left(-\frac{\beta^2 \rho^2}{2}\right)} &\mbox{ if $\rho<\mu$} \, ,\\
 A_2 \exp{\left(-\frac{\eta \rho}{2}\right)} &\mbox{ if $\rho > \mu$}\, .
       \end{array} \right.
\label{eq:MV}
\end{equation}     
This {\it ansatz} involves five parameters, three of which ($A_1$, $A_2$, and $\eta$) are obtained from the boundary conditions and normalization, and we are left with two variational parameters ($\beta$, and $\mu$). We note that $\eta$ is obtained from $\eta = 2 \beta ^2 \mu$. The adopted matched variational wavefunction gives an exchange coupling which fits well with the asymptotic values given by more rigorous calculations~\cite{ponomarev992} and which is orders of magnitude larger than the one we previously calculated using a pure gaussian variational form for the electron wavefunction at the interface.~\cite{calderon07} The exchange we are getting is of the order of the one estimated for gated quantum dots in GaAs.~\cite{laird06}

\section{Discussion and conclusion}
\label{sec:discuss}
We have discussed here some of the relevant parameters that would allow  elementary operations involved in  the manipulation of donor electrons in Si close to a SiO$_2$ interface for quantum computing applications. For our calculations we have considered an ideal semi-infinite SiO$_2$ barrier. However, in real devices~\cite{kenton06,lansbergen08,morello} metallic gates are located on top of a relatively thin (few nm wide) SiO$_2$ barrier. It seems then that a more realistic description of the devices should take into account the metallic gates. There exists a calculation similar to ours in the opposite limit of an infinitesimal SiO$_2$ barrier separating the Si layer from the metallic gate.~\cite{slachmuylders08} In that case, $Q=1$ and the image charges have opposite sign to the real charges. In particular, the donor image potential will be repulsive instead of attractive and this can seriously deteriorate the confinement of the electrons along the xy-plane at the interface. For a realistic case of a SiO$_2$ barrier width of $\sim 2$ nm and an interface-donor distance $d=4-5$ nm,~\cite{lansbergen08} we would be in between the two extreme cases of a semi-infinite SiO$_2$ barrier considered here and an infinitesimal one considered in Ref.~\onlinecite{slachmuylders08}. In particular, the width of the wavefunction at the interface $1/\beta$ would be larger than reported here. By comparing our $1/\beta$ with the results reported in Ref.~\onlinecite{slachmuylders08} [see $\sqrt{\rho^2}/a_0$ in the inset of Fig. 4] we can estimate at most a doubling of $1/\beta$ when metallic gates are included. This pushes the upper limit for the donor planar density towards smaller values, still keeping our qualitative conclusions valid. 

Another relevant issue is related to the modification of the spin coherence time T$_2$ when donors are located close to a Si surface. Different experiments \cite{schenkel06,huebl08} have revealed that T$_2$ gets smaller than in bulk  when donors are located close to an interface: bulk values reach $T_2^{\rm bulk} \sim 60$ ms while close to a surface they range from $T_2^{\rm surf}\sim 0.1-1$ ms~\cite{schenkel06} to $T_2^{\rm surf}\sim 1\,\mu s$~\cite{huebl08} for different samples. T$_2$ depends on the quality of the interface (it is longer for a hydrogen passivated surface than for a Si/SiO$_2$ interface~\cite{schenkel06}) and the distance from the donor to the interface $d$. Some effects that may be responsible for the decrease of coherence times are the magnetic noise produced by impurities and other defects (like P$_b$ centers) at the interface,~\cite{desousa07} and/or the recombination of donor electrons on the P$_b$ centers.~\cite{huebl08} These observations imply that increasing the interface quality may be crucial for the practical implementation of a Si-based quantum computer.  

In summary, we have described the basic conditions that would allow the manipulation of donor electrons in Si by external fields, taking into account the multivalley structure of the conduction band of Si. Experiments are getting close to the limit of isolated donors~\cite{lansbergen08,morello} where the  operations discussed here are relevant. In fact, qualitative agreement for the behavior of $F_c$ versus $d$ has already been found in samples with isolated donors in a Si FinFET,~\cite{lansbergen08} in a situation where the donor is closer to the interface ~\cite{koillerN&V} than we considered in the present study. Both situations (small and large values of $d$ as compared to the effective electronic confinement lengths) are important, and should be carefully investigated in the context of applications of donor electrons as carriers of quantum information through its charge or spin degrees of freedom.

\begin{acknowledgments}
This work is supported by LPS and NSA. M.J.C.  acknowledges support from Ram\'on y Cajal Program and MAT2006-03741 (MICINN, Spain). B.K. also acknowledges support from CNPq, FUJB, Millenium Institute
on Nanotechnology - MCT, and FAPERJ.
\end{acknowledgments}

\end{document}